\documentclass[noshowpacs,amsmath,
twocolumn,
superscriptaddress,
8pt,
aps,prb]{revtex4-2}
\usepackage{setspace}
\usepackage{amsmath}
\usepackage{multirow}
\usepackage{graphicx}

\usepackage{verbatim}
\usepackage{amsfonts}
\usepackage{amssymb}
\usepackage{epstopdf}
\usepackage{dcolumn}
\usepackage{xcolor}
\usepackage{verbatim}
\usepackage{hyperref}
\usepackage{siunitx}
\usepackage[version=4]{mhchem}
\usepackage{textcomp}
\usepackage[left]{lineno}
\setcitestyle{super} 
\DeclareGraphicsExtensions{.pdf,.eps,.png,.jpg,.mps}

\begin{document}

\preprint{APS/123-QED}

\title{Integrated soliton microcombs beyond the turnkey limit}

\author{Ze Wang$^{1,*}$, Tianyu Xu$^{1,*}$, Yuanlei Wang$^{1,2*}$, Kaixuan Zhu$^{1,*}$, Xinrui Luo$^{1,}$, Haoyang Luo$^{1}$, Junqi Wang$^{1}$, Bo Ni$^{1,3}$, Yiwen Yang$^{1}$, Qihuang Gong$^{1,3,4}$, Yun-Feng Xiao$^{1,3,4}$, Bei-Bei Li$^{2}$, and Qi-Fan Yang$^{1,3,4\dagger}$\\
$^1$State Key Laboratory for Artificial Microstructure and Mesoscopic Physics and Frontiers Science Center for Nano-optoelectronics, School of Physics, Peking University, Beijing 100871, China\\
$^2$Beijing National Laboratory for Condensed Matter Physics, Institute of Physics, Chinese Academy of Sciences, Beijing 100190, China\\
$^3$Peking University Yangtze Delta Institute of Optoelectronics, Nantong, Jiangsu 226010, China\\
$^4$Collaborative Innovation Center of Extreme Optics, Shanxi University, Taiyuan 030006, China\\
$^{*}$These authors contributed equally to this work.\\
$^{\dagger}$Corresponding author: leonardoyoung@pku.edu.cn}

\begin{abstract}
Soliton microcombs generated in optical microresonators are accelerating the transition of optical frequency combs from laboratory instruments to industrial platforms. Self-injection locking (SIL) enables direct driving of soliton microcombs by integrated lasers, providing turnkey initiation and improved coherence, but it also pins the pump close to resonance, limiting both spectral span and tuning flexibility. Here we theoretically and experimentally demonstrate that introducing a thermally tunable auxiliary microresonator extends the bandwidth of SIL soliton microcombs. By engineering hybridization of the pumped resonance, we achieve deterministic access to single-soliton states and then push operation into a far-detuned regime inaccessible to direct initiation. The resulting combs reach a near-200-nm span at a 25-GHz repetition rate, while preserving the SIL-enabled noise suppression throughout. Moreover, the added degree of freedom afforded by the coupled-resonator architecture enables orthogonal control of the comb’s repetition rate and center frequency. These advances expand the spectral reach and controllability of integrated soliton microcombs for information processing and precision metrology.
\end{abstract}

\maketitle 

Interest in deploying optical frequency combs in practical settings—spanning metrology, communications, and computing—is rapidly growing \cite{diddams2020optical}. Kerr microcombs, generated in high-$Q$ optical microresonators via Kerr-driven four-wave mixing, represent a leading route to this goal \cite{Kippenberg2018}. A particularly important class of Kerr microcombs is the soliton microcomb, in which dissipative Kerr solitons yield femtosecond pulses and phase-locked, equally spaced comb lines with a well-defined spectral envelope \cite{herr2014temporal,yi2015soliton,brasch2016photonic,gong2018high,he2019self}. Such sources can provide bidirectional optical–microwave links for optical clocks \cite{newman2019architecture}, pulse trains for precision ranging \cite{suh2018soliton,trocha2018ultrafast,jang2021nanometric}, and multiwavelength carriers for high-speed data transmission \cite{marin2017microresonator,jorgensen2022petabit}. While initial demonstrations of soliton microcombs employed simple single-ring microresonators, their performance, particularly in terms of power efficiency and spectral span, has been substantially enhanced by incorporating more advanced resonator geometries and coupling schemes~\cite{obrzud2017temporal,bao2019laser,yu2021spontaneous,helgason2023surpassing,ulanovsynthetic2024,zhu2025ultra}.

\begin{figure*}[htbp]
  \centering
  \includegraphics[width=\linewidth]{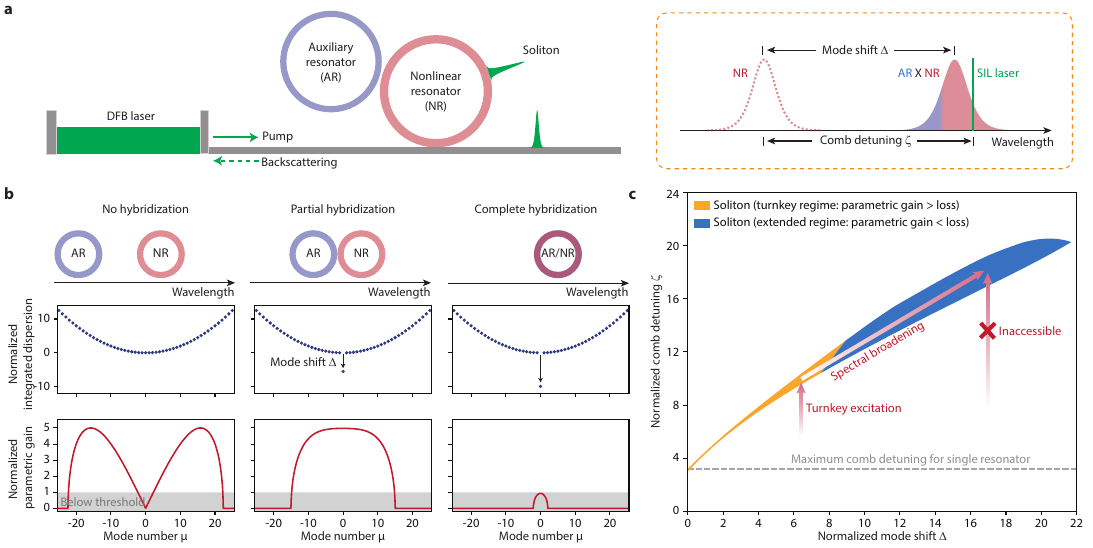}
  \caption{\textbf{Dynamics of self-injection locking (SIL) in coupled microresonators.}
  \textbf{a,} Schematic of a coupled-resonator device directly pumped by a distributed-feedback (DFB) laser. The original NR mode is indicated by dashed lines, and the NR (red) and AR (blue) fractions of the hybridized mode are encoded in color. 
  \textbf{b,} Effect of pump-mode hybridization on the NR integrated dispersion (middle) and on the parametric gain spectrum (bottom). Both quantities are normalized to half the NR linewidth, $\kappa_\mathrm{NR}/2$.
  \textbf{c,} Phase diagram of soliton existence under SIL conditions, plotted as a function of comb detuning and mode shift. Arrows indicate the pathway from turnkey excitation to spectrally broadened soliton microcombs.}
  \label{fig:fig1}
\end{figure*}

Photonic integration has recently transformed soliton microcombs: by directly coupling on-chip lasers or gain chips to microresonators, soliton states can form autonomously via self-injection locking (SIL)~\cite{liang2015high,pavlov2018narrow,stern2018battery,raja2019electrically,shen2020integrated,voloshin2021dynamics,jin2021hertz,ulanovsynthetic2024,wang2025compact}. Beyond turnkey operation, SIL narrows the pump-laser linewidth and allows the use of inexpensive, low-coherence lasers, enabling compact, mass-producible comb sources suitable for deployment outside the laboratory~\cite{wang2025compact}. However, SIL also introduces two key constraints: (i) the locking mechanism limits the achievable pump–resonator detuning, even though large detuning is essential for broadband spectra; and (ii) locking removes an independent tuning degree of freedom, whereas orthogonal control of the repetition rate and comb center frequency requires two. Photonic-crystal microresonators can partially alleviate these limitations by shifting the pump mode off the native dispersion curve, thereby permitting larger detuning while maintaining lock. In practice, however, turnkey comb initiation becomes increasingly difficult as the pump-mode shift grows, ultimately limiting the attainable detuning~\cite{ulanovsynthetic2024}.

Here, we overcome this limitation by introducing a reconfigurable coupled-resonators architecture for integrated soliton microcombs. Although similar structures have previously been used to enhance soliton efficiency~\cite{helgason2023surpassing}, this is, to our knowledge, their first application in the SIL regime. By dynamically adjusting the pump-mode frequency via an auxiliary resonator, an initiation–tuning protocol grants access to detuning values far beyond the turnkey SIL range, where markedly broader combs emerge. We experimentally implement this approach using a distributed-feedback (DFB) laser pumping Si$_3$N$_4$ coupled microresonators and achieve a near-200-nm span at a 25~GHz repetition rate, while preserving SIL-enabled noise suppression throughout. Moreover, the auxiliary resonator provides an additional control degree of freedom that we exploit to realize orthogonal tuning of the repetition rate and pump frequency of the comb.

\begin{figure*}[htbp]
  \centering
  \includegraphics[width=\linewidth]{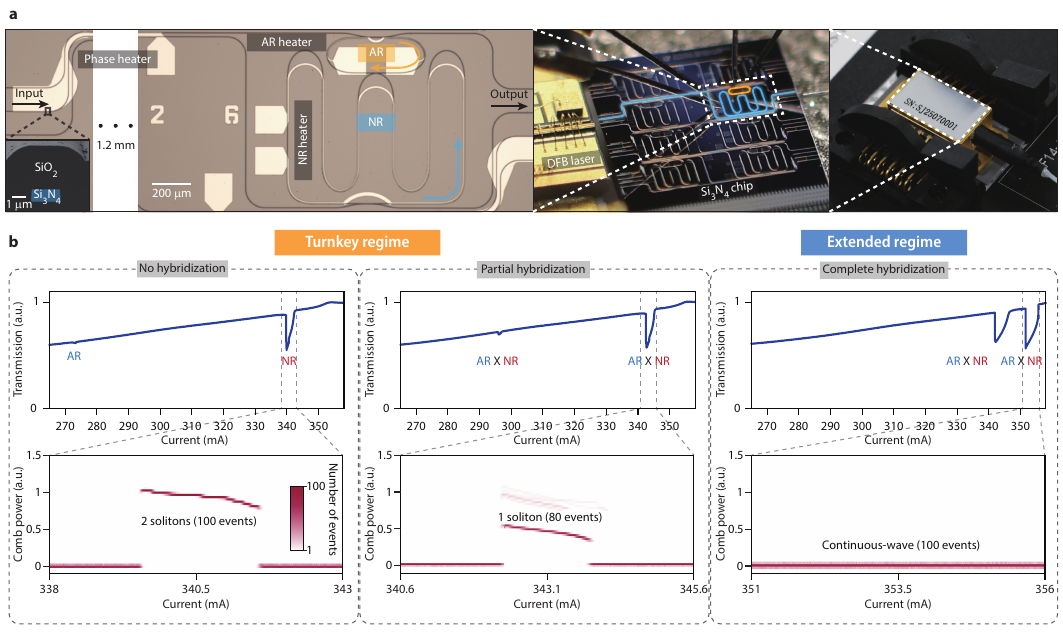}
  \caption{\textbf{Generation of soliton microcombs.}
  \textbf{a,} Device photographs. From left to right: Si$_3$N$_4$ photonic chip, DFB laser directly butt-coupled to the chip, and the fully packaged assembly in a butterfly package.
  \textbf{b,} Upper panel: measured transmission spectra at different hybridization levels, obtained by varying the AR heater power. The AR resonance, NR resonance, and hybridized resonances (AR~$\times$~NR) are indicated. Lower panel: comb power versus DFB drive current over 100 consecutive scans.}
  \label{fig:fig2}
\end{figure*}

\begin{figure*}[t]
  \centering
  \includegraphics[width=\linewidth]{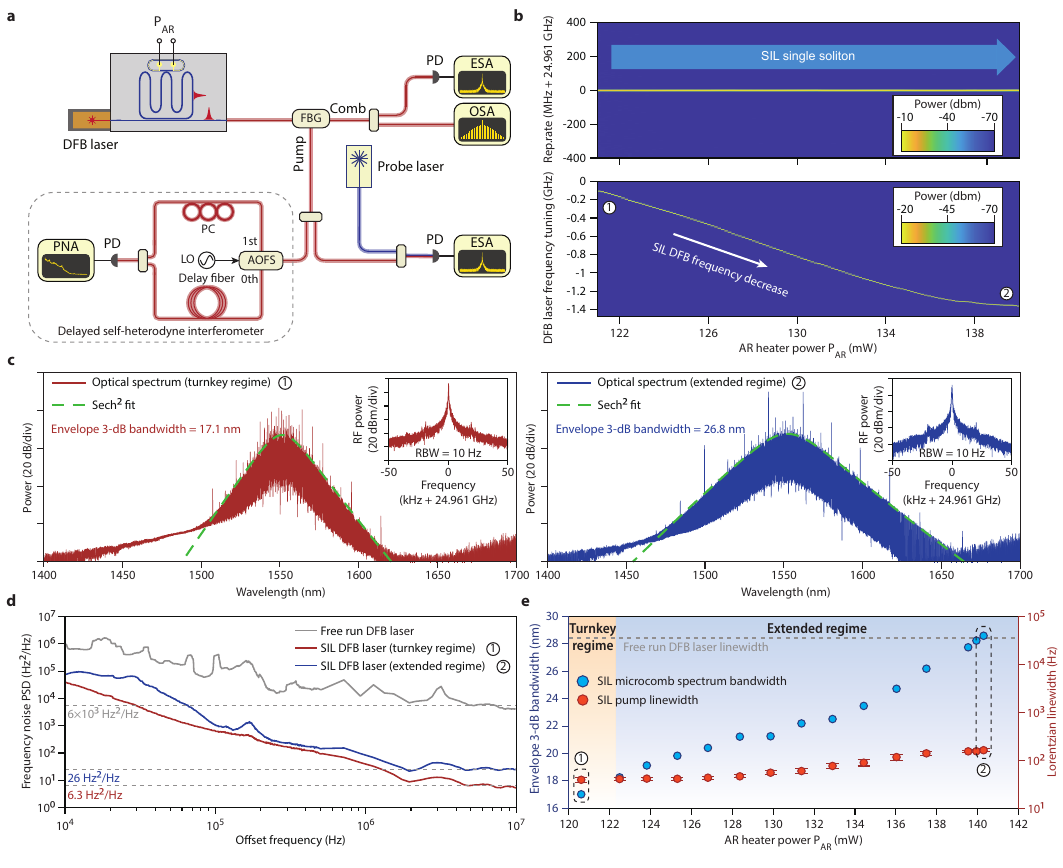}
  \caption{\textbf{Tuning and characterization of soliton microcombs.}
  \textbf{a,} Experimental setup. PD: photodetector; ESA: electrical spectrum analyzer; OSA: optical spectrum analyzer; FBG: fiber Bragg grating; PNA: phase-noise analyzer; LO: local oscillator; PC: polarization controller; AOFS: acousto-optic frequency shifters.
  \textbf{b,} Measured comb electrical beat note (upper panel) and DFB laser frequency (lower panel) as a function of AR heater power.
  \textbf{c,} Optical spectra of soliton microcombs in the turnkey (left) and extended (right) regimes at the operating points indicated in \textbf{b}. The 3-dB spectral bandwidth is extracted by fitting the spectra with $\mathrm{sech}^2$ envelopes. Insets: corresponding electrical beat notes of the combs.
  \textbf{d,} Power spectral density (PSD) of the frequency noise of the free-running and SIL DFB lasers. The noise floor at high offset frequencies is indicated.
  \textbf{e,} Measured 3-dB spectral-envelope bandwidth (left axis) and Lorentzian linewidth (right axis) of the microcomb as a function of AR heater power.}
  \label{fig:fig3}
\end{figure*}

\begin{figure}[t]
  \centering
  \includegraphics[width=\linewidth]{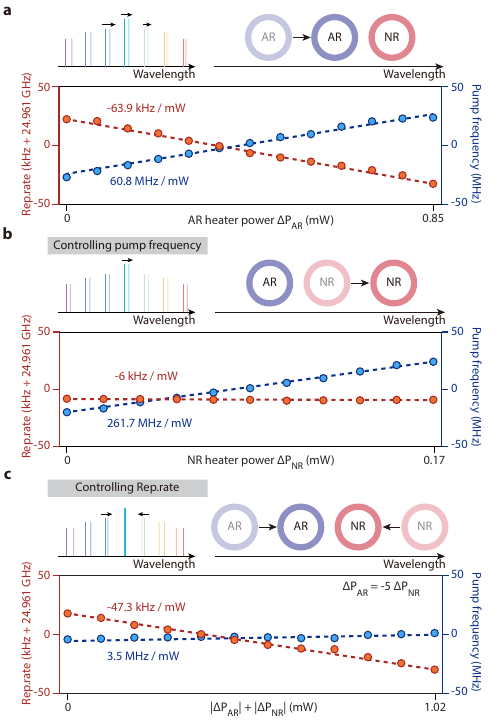}
  \caption{\textbf{Orthogonal control of comb properties.}
\textbf{a,} Repetition rate (red) and pump frequency (blue) via AR tuning.
\textbf{b,} Repetition rate (red) and pump frequency (blue) via NR tuning.
\textbf{c,} Repetition rate (red) and pump frequency (blue) via combined AR and NR tuning.
  }
  \label{fig:fig4}
\end{figure}

\vspace{6pt}
\noindent\textbf{Results}

\noindent
\textbf{Principle}

\noindent Figure~\ref{fig:fig1}a illustrates the architecture, in which a nonlinear resonator (NR) is coupled to a bus waveguide and side-coupled to an auxiliary resonator (AR). A DFB laser is directly coupled to the bus waveguide without an optical isolator. Fabrication-induced backscattering provides optical feedback that drives SIL~\cite{gorodetsky2000rayleigh,liang2015ultralow,kondratiev2017self,stern2017compact,li2021reaching,jin2021hertz,shen2024reliable}, and the feedback phase—the phase of the reinjected field with respect to the emitted pump—is tuned to optimize locking. At elevated pump powers, Kerr nonlinearity lifts the degeneracy between the forward- and backward-propagating modes, shifting the optimal SIL point slightly to the red of the forward resonance. This red detuning satisfies the soliton existence condition and enables turnkey comb initiation~\cite{liang2015high,pavlov2018narrow,shen2020integrated,voloshin2021dynamics,jin2021hertz}, but the resulting nonlinear splitting is small, limiting the accessible detuning and comb span.

Coupling between the AR and NR induces mode hybridization, producing a hybridized resonance that is red-shifted from the NR pump resonance by an amount $\Delta$. SIL then locks the DFB laser to this hybridized resonance, yielding a larger comb detuning $\zeta$ defined with respect to the unperturbed NR mode family; to leading order, the spectral span scales as $\sqrt{\zeta}$. The hybridization is described by the NR integrated dispersion (without Kerr), $D_{\mathrm{int}}(\mu)=\omega_\mu-\omega_0-D_1\mu-\Delta\,\delta_{\mu 0}\approx D_2\mu^2/2-\Delta\,\delta_{\mu 0}$, where $\mu$ indexes modes relative to the pump, $\omega_{\mu}/2\pi$ and $\omega_{0}/2\pi$ denote the $\mu$th and pump resonance frequencies and $D_{1}/2\pi$ represents the free spectral ranges (FSR) of the NR. $\Delta$ is the frequency shift applied to the pump mode ($\mu=0$) of NR, $\delta_{\mu0}$ is the Kronecker delta function (\(\delta_{00}=1\)), and anomalous group velocity dispersion ($D_2>0$) is required for soliton formation. As the AR resonance is tuned through the NR pump mode, the hybridized mode is shifted from the original frequency, with $\Delta$ maximized at exact alignment (Fig.~\ref{fig:fig1}b). This mechanism provides continuous control of $\Delta$ and hence of $\zeta$.

Comb initiation occurs when the parametric gain exceeds the resonator loss. Increasing $\Delta$ pulls the parametric gain lobes toward $\mu=0$, merges them, and eventually suppresses the peak gain below threshold (Fig.~\ref{fig:fig1}b). To quantify the resulting behavior, we construct a phase diagram in the $(\zeta,\Delta)$ plane at a pump power five times the parametric-oscillation threshold of the standalone NR, subject to the SIL constraints (see Methods). We identify soliton states by scanning the feedback phase and the laser free-running frequency. Without hybridization ($\Delta=0$), the accessible comb-detuning range is narrow, with a maximum normalized detuning of 3.2 (in units of $\kappa_{\mathrm{NR}}/2$). Increasing $\Delta$ broadens the soliton region and extends the attainable detuning to 12.7 at $\Delta = 8.8$. For $\Delta<8.8$, the parametric gain still exceeds the resonator loss, so solitons form automatically when the system is tuned to a preset operating point; we refer to this as the turnkey regime~\cite{shen2020integrated}. Beyond this point, the system enters an extended regime in which direct initiation is suppressed because the parametric gain falls below the resonator loss, yet solitons seeded in the turnkey regime persist up to $\Delta = 19.7$, with detuning as large as 20.6 (see Supplementary Information). Access to these larger detunings substantially broadens the comb span and relies on a continuously tunable mode shift, in contrast to photonic-crystal ring microresonators, where $\Delta$ is fixed~\cite{ulanovsynthetic2024}.

\vspace{6pt}
\noindent
\textbf{Comb generation in the turnkey regime}

\noindent The device, including the coupled microresonators and the DFB laser, is shown in Fig. \ref{fig:fig2}a. The microresonators are fabricated via a subtractive process on a 786-nm-thick Si$_3$N$_4$-on-insulator wafer (see Methods). The NR and AR exhibit FSRs of 25 GHz and 115 GHz, respectively, and the resulting FSR mismatch ensures that only the pumped NR mode hybridizes with an AR mode near the pump wavelength. The NR, with a waveguide width of 1.8 $\mu$m, has an intrinsic quality factor $Q_0 = 7.43\times 10^6$ and a coupling quality factor $Q\mathrm{e} = 4.64\times 10^6$. The AR, with a width of 1.3 $\mu$m, exhibits $Q_0 = 1.15\times 10^6$ and $Q\mathrm{e} = 9.09\times 10^6$. The gap between the NR and AR is 200 nm, corresponding to an inter-resonator coupling rate $G/2\pi = 1.05$ GHz. Integrated heaters on the microresonators provide independent thermal tuning of each resonator, and a phase heater on the bus waveguide controls the feedback phase. A commercial DFB laser is directly coupled to the Si$_3$N$_4$ chip and, after correcting for insertion loss, delivers approximately 35 mW of optical power into the bus waveguide. To enhance stability, the DFB laser and coupled microresonators are co-packaged in a standard butterfly module, which preserves phase-stable optical feedback and enables portable operation outside the laboratory.

Comb generation across three hybridization regimes is experimentally verified in Fig.~\ref{fig:fig2}b. For each heater setting, we record the transmission and comb power as functions of the DFB current, which, in the absence of SIL, sets the free-running laser frequency. Without hybridization, scanning the laser across the NR resonance predominantly yields a two-soliton state, because the maximum parametric gain occurs at multiple free-spectral-range spacings from the pump (see Supplementary Information). Nevertheless, turnkey comb generation remains possible by directly biasing the laser current to the corresponding operating point (for example, 340.5 mA). Next, the AR is tuned closer to the NR to realize partial hybridization, such that one hybridized resonance is red-shifted relative to the original NR resonance. In this condition, ramping up the laser current most frequently produces single-soliton states, providing a favorable operating point for deterministic turnkey initiation. This behavior arises because the maximum parametric gain is shifted closer to the pump and the first generated sidebands appear at mode numbers $\mu=\pm 1$ (Fig.~\ref{fig:fig1}b). When the AR is tuned even closer to the NR, the system approaches full hybridization, and the transmission spectrum exhibits two hybridized resonances with comparable depth. Pumping the most red-shifted hybrid mode then results in no comb formation, indicating that the system has entered the extended regime.

\vspace{6pt}
\noindent
\textbf{Comb broadening in the extended regime}

\noindent As outlined in Fig.~\ref{fig:fig3}a, we comprehensively characterize the SIL soliton state: the optical spectrum is measured with an optical spectrum analyzer, the repetition rate is obtained from the photocurrent of a fast photodetector on an electrical spectrum analyzer, the pump frequency is tracked by heterodyning the filtered DFB output with a fixed probe laser, and the DFB frequency noise is measured using a delayed self-heterodyne interferometer and a phase-noise analyzer. The AR heater power $P_{\mathrm{AR}}$ parametrizes the AR–NR hybridization and sets the NR mode shift. A single soliton is initiated in the turnkey regime at $P_{\mathrm{AR}} \approx 120.5$~mW and subsequently driven into the extended regime by ramping up $P_{\mathrm{AR}}$. The resulting mode shift causes the DFB frequency to decrease approximately linearly due to SIL (Fig.~\ref{fig:fig3}b), with a gradual roll-off near the end of the tuning range indicating the approach to the maximal mode shift. In both the turnkey and extended regimes, the SIL microcomb remains a stable single soliton, as evidenced by a single, narrow repetition-rate line at 24.961~GHz. Meanwhile, the soliton spectrum broadens markedly: the 3-dB envelope bandwidth increases from 17.1~nm (turnkey) to 26.8~nm (extended), and the nonlinear conversion efficiency increases from 4.4\% to 7.9\% (Fig.~\ref{fig:fig3}c). Spectral protrusions relative to the $\mathrm{sech}^2$ envelope mostly arise from periodic mode hybridization induced by the interleaved AR and NR free spectral ranges, which perturb the integrated dispersion of NR. In the extended regime, the larger mode shift and increased soliton detuning reduce the impact of this distortion and smooth the central spectrum. Further engineering of the AR free spectral range to suppress periodic hybridization is expected to yield a spectrum that more closely follows a pure $\mathrm{sech}^2$ envelope.

A key advantage of SIL is the reduction of pump frequency noise. The achievable noise suppression depends on the microresonator quality factor and the strength of backscattering, which typically decreases at large detuning as intracavity power is reduced \cite{jin2021hertz}. In the coupled-resonators architecture, however, laser injection into the hybridized resonance remains efficient even at large comb detuning, so the intracavity power and associated backscattering remain sufficient to sustain noise reduction. We therefore measure the pump frequency noise in both the turnkey and extended regimes (Fig.~\ref{fig:fig3}d), comparing their spectra at high offset frequencies ($f \gtrsim 10~\mathrm{kHz}$) to suppress the influence of environmental and thermal noise. Within the extended regime, SIL-enabled noise suppression remains largely intact even at the maximal mode shift, with only a slight increase in the DFB frequency-noise power spectral density relative to the turnkey regime (Fig.~\ref{fig:fig3}d). This small increase in SIL-locked laser noise is attributable to enhanced AR–NR hybridization: with $Q_{\mathrm{NR}} \sim 4 \times 10^{6}$ and $Q_{\mathrm{AR}} \sim 1 \times 10^{6}$, stronger coupling lowers the $Q$ of the predominantly NR-like hybrid mode and slightly weakens linewidth compression. Figure~\ref{fig:fig3}e summarizes the comb properties over the entire tuning range: the 3-dB spectral-envelope bandwidth increases monotonically with $P_{\mathrm{AR}}$, while the Lorentzian linewidth, inferred from the high-frequency noise spectrum (see Methods), remains strongly suppressed relative to the free-running DFB, maintaining $>20~\mathrm{dB}$ reduction.

\vspace{6pt}
\noindent
\textbf{Orthogonal control of comb properties}

\noindent The coupled-microresonator architecture enables two-degree-of-freedom control of the soliton repetition rate and pump frequency by exploiting their distinct sensitivities to thermal tuning of the AR and NR. Tuning the AR modifies both the mode shift, and hence the pump frequency, and the comb detuning, thereby changing the repetition rate via Raman self-frequency shifts~\cite{karpov2016raman,yi2016theory,yang2016spatial} and dispersive-wave recoil~\cite{brasch2016photonic,yi2017single}. In contrast, tuning the NR predominantly shifts the pump resonance and thus the pump frequency, while only weakly perturbing the comb detuning and repetition rate; moreover, the opposing contributions from the slight change in comb detuning and the change in the FSR of the NR partially cancel. By forming appropriate linear combinations of AR and NR heater powers, these two control channels can be recombined into orthogonal eigenmodes that independently address either the repetition rate or the pump frequency.

Experimentally, we operate in the extended detuning regime with AR and NR heater powers of 127~mW and 1~mW, respectively. Tuning the AR yields an approximately linear decrease in the repetition rate of $-63.9$~kHz/mW and a concomitant increase in the pump frequency of 60.8~MHz/mW (Fig.~\ref{fig:fig4}a). Near this operating point, tuning only the NR produces a tenfold smaller change in repetition rate ($-6$~kHz/mW) but a much larger pump-frequency shift of 261.7~MHz/mW (Fig.~\ref{fig:fig4}b). Based on these response coefficients, we implement orthogonal control by tuning the AR and NR in opposite directions with an appropriate $\sim 5{:}1$ ratio of heater-power changes (Fig.~\ref{fig:fig4}c). This strategy yields an effective repetition-rate tuning coefficient of $-47.3$~kHz/mW (normalized to $|\Delta P_\mathrm{AR}|+|\Delta P_\mathrm{NR}|$), while suppressing the pump-frequency tuning to only 3.5~MHz/mW. Although the orthogonality can be further refined to reduce residual cross-talk, the demonstrated control is already sufficient to enable potential full stabilization of the soliton microcomb~\cite{wildi2024phase,liu2025separable}.

\vspace{6pt}
\noindent
\textbf{Discussion}

\noindent In conclusion, we have introduced an integrated soliton microcomb architecture that operates beyond the conventional turnkey limit while preserving the simplicity and coherence benefits of SIL. Although demonstrated here in Si$_3$N$_4$ microresonators, the concept is generic and can be extended to other Kerr-comb platforms, including LiNbO$_3$~\cite{he2019self,song2025stable,nie2025soliton}, LiTaO$_3$~\cite{wang2024lithium}, AlN~\cite{gong2018high}, and SiON~\cite{wang2018robust}, among others. Several of these platforms offer intrinsic electro-optic or piezoelectric tuning, enabling fast, low-power, and programmable comb control. Even in platforms without native electro-optic functionality, heterogeneously integrated piezoelectric actuators can substantially increase the tuning speed relative to thermal heaters~\cite{liu2020monolithic,liu2025separable}. These capabilities are expected to further enhance locking performance in applications such as optical frequency division~\cite{kudelin2023photonic,sun2023integrated,jin2025microresonator,ji2025dispersive,sun2025microcavity}, optical clocks~\cite{newman2019architecture}, and optical frequency synthesis~\cite{spencer2018optical}.

The increased spectral span and high level of integration make this platform directly relevant to a broad range of applications. In optical frequency division, the larger comb bandwidth supports higher division ratios and therefore lower phase noise in the generated microwaves~\cite{kudelin2023photonic,sun2023integrated,jin2025microresonator,ji2025dispersive,sun2025microcavity}. In dual-comb spectroscopy, the broader spectral coverage enables simultaneous interrogation of more molecular species~\cite{suh2016microresonator,dutt2018chip,yang2019vernier}. If the NR free spectral range is increased to the few-hundred-gigahertz regime, the resulting comb can provide line spacings compatible with current wavelength-division-multiplexing formats for high-capacity telecommunications~\cite{marin2017microresonator,jorgensen2022petabit}. Moreover, the NR dispersion and $Q$ factor can be engineered to realize either broadband or high-power combs, tailored to specific system requirements. Overall, the presented architecture represents a significant advance over conventional SIL soliton microcombs and is well positioned for deployment across a wide range of next-generation coherent photonic systems.

\bigskip

\noindent\textbf{Methods}
\begin{footnotesize}

\medskip
\noindent{\bf Phase diagram.}
Reconfigurable pump-mode hybridization induced by the AR is modeled as an effective frequency shift of the pump mode within the NR mode family~\cite{helgason2023surpassing}. To obtain the phase diagram in Fig.~\ref{fig:fig1}c, we model the SIL system with pump-mode hybridization in terms of the soliton field $\psi_{\mathrm{S}}$, the backscattered field $\psi_{\mathrm{B}}$, and the dynamics of the SIL pump laser. Their evolution is governed by the following equations of motion~\cite{godey2014stability,kondratiev2017self,yu2021spontaneous}:
\begin{equation}
\label{eq:LLEtot}
\begin{aligned}
    \frac{\partial \psi_{\mathrm{S}}}{\partial \tau}&=-(1 + i\zeta)\psi_{\mathrm{S}}+i d_{2}\frac{\partial^2 \psi_{\mathrm{S}}}{\partial \phi^2}+i(|\psi_\mathrm{S}|^2+2|\psi_\mathrm{B}|^2)\psi_\mathrm{S}+i\beta\psi_\mathrm{B} \\
    &\hphantom{{}=} + i\Delta\overline{\psi_\mathrm{S}}+ f ,\\[6pt]
    \frac{\partial \psi_{\mathrm{B}}}{\partial \tau}&=-(1 + i\xi)\psi_{\mathrm{B}}+i(|\psi_{\mathrm{B}}|^2+2P)\psi_{\mathrm{B}}+i\beta\overline{\psi_\mathrm{S}},\\[6pt]
    \xi&=\Delta_\mathrm{L}+K_0\mathrm{Im}\left[(1-i\alpha_\mathrm{g})e^{i2\phi_\mathrm{B}}\frac{\psi_\mathrm{B}}{f}\right].
\end{aligned}
\end{equation}
\noindent
Here, $\tau$ is the normalized slow time and $\phi$ is the angular coordinate in the co-rotating frame; $\zeta$, $\xi$, and $\Delta_{\mathrm{L}}$ are the normalized comb, pump, and laser detunings (in units of $\kappa_{\mathrm{NR}}/2$), and the normalized mode shift $\Delta$ links pump and comb detunings via $\zeta=\xi+\Delta$. The parameter $d_2$ denotes the normalized group-velocity dispersion of the NR; $\beta$ is the dimensionless backscattering coefficient (in units of $\kappa_{\mathrm{NR}}/2$); and $|f|^{2}$ is the normalized pump power, expressed in units of the parametric-oscillation threshold of the standalone NR. $K_{0}$ is the locking bandwidth, $\alpha_\mathrm{g}$ the linewidth–enhancement factor, and $\phi_{B}$ the feedback phase (i.e., the phase delay between the NR and the laser). We define $P=\overline{|\psi_{\mathrm{S}}|^{2}}$ as the normalized total intracavity power and $\overline{\psi_\mathrm{S}}$ as the normalized amplitude of the NR pump mode. For a given $\Delta$, we sweep $\Delta_{\mathrm{L}}$ and $\phi_{\mathrm{B}}$ to determine the range of comb detunings that support solitons, holding all other parameters fixed at $|f|^2=5$, $\beta=0.1$, $K_0=50$, $\alpha_\mathrm{g}=0$, and $d_2=0.02$. To identify solitons in the turnkey regime, quantum noise corresponding to one-half photon per mode is injected into $\psi_{\mathrm{S}}$; a turnkey soliton obtained in this way is then used as the initial condition to map soliton persistence in the extended regime. A detailed derivation and simulation procedure are provided in the Supplementary Information.

\medskip
\noindent{\bf Device fabrication.}
The Si$_3$N$_4$ coupled microresonators are realized on a 4-inch wafer using a subtractive fabrication process~\cite{wang2025compact}. First, a 786~nm-thick Si$_3$N$_4$ layer is deposited in two steps onto a wet-oxidized silicon substrate that incorporates stress-relief structures. The waveguide layout is then defined by electron-beam lithography, and the pattern is transferred into the Si$_3$N$_4$ film by dry etching. To eliminate residual N–H and Si–H bonds, the wafer is annealed at 1200$^{\circ}$C. A SiO$_2$ cladding layer is subsequently deposited and densified in a second annealing step. Metal microheaters are patterned using a lift-off process. Finally, the wafer is diced into $5~\mathrm{mm} \times 5~\mathrm{mm}$ chips.

\medskip
\noindent{\bf Laser linewidth measurement.}
The pump-laser frequency noise is measured with a delayed self-heterodyne interferometer. In this scheme, the laser output is divided into two paths: one path is frequency shifted by 55~MHz using an acousto-optic modulator, while the other is sent through a 1-km optical-fiber delay line. The two paths are then recombined, detected on a photodiode, and the resulting beat signal is analyzed with a phase-noise analyzer to obtain the single-sideband phase-noise spectrum $S_\mathrm{\phi}(f)$. The corresponding frequency-noise PSD $S_\mathrm{F}(f)$ is obtained via
\begin{equation}
\label{eq:DSH}
\begin{aligned}
S_\mathrm{F}(f) &= S_\mathrm{\phi}(f)\frac{f^2}{4\sin^2{(\pi f\tau_d)}},
\end{aligned}
\end{equation}
\noindent
where $\tau_d$ denotes the interferometric delay time (4.8821~\textmu s in this work). Because the denominator vanishes at offset frequencies satisfying $\sin{(\pi f\tau_d)}=0$, the analysis in Fig.~\ref{fig:fig3}e is restricted to offset frequencies $(N+1/2)/\tau_d$, with $N$ a positive integer. The Lorentzian linewidth reported in Fig.~\ref{fig:fig3}e is then extracted from the high-frequency plateau of $S_{\mathrm{F}}(f)$ as
\begin{equation}
\label{eq:lorentz}
\begin{aligned}
\delta\nu &= 2\pi\sum_{N=20}^{59}\frac{S_\mathrm{F}\left(\frac{N+1/2}{\tau_d}\right)}{30},
\end{aligned}
\end{equation}
\noindent
and the error bars in Fig.~\ref{fig:fig3}e correspond to the standard deviation of $S_{\mathrm{F}}(f)$ over the discrete offset frequencies used in Eq.~\eqref{eq:lorentz}.

\medskip
\noindent{\bf Feedback phase optimization.}
When the AR is thermally tuned in the extended regime, the feedback phase is perturbed by thermal crosstalk, which can degrade the noise-reduction performance of self-injection locking. For the Lorentzian linewidth data in Fig.~\ref{fig:fig3}e, we therefore re-optimize the voltage applied to the phase heater ($E_\mathrm{phase}$) after each AR adjustment to restore the optimal feedback phase. Specifically, we perform a fine sweep of $E_\mathrm{phase}$ over a narrow range with the smallest available step size, compute the corresponding Lorentzian linewidth at each setting, and choose the minimum as the operating point.

\end{footnotesize}

\bibliography{PRL/ref.bib}

\medskip
\noindent\textbf{Acknowledgments}

\begin{footnotesize}
\noindent The authors thank Yinke Cheng, Xin Zhou, Jincheng Li, Zhigang Hu for assistance in fabrication, and Du Qian for assistance on sample photography. This work was supported by National Key R\&D Plan of China (Grant No. 2023YFB2806702), National Natural Science Foundation of China (12293050), and the High-performance Computing Platform of Peking University. The fabrication in this work was supported by the Peking University Nano-Optoelectronic Fabrication Center, Micro/nano Fabrication Laboratory of Synergetic Extreme Condition User Facility (SECUF), Songshan Lake Materials Laboratory, and the Advanced Photonics Integrated Center of Peking University.
\end{footnotesize}
\medskip

\noindent\textbf{Author contributions}

\begin{footnotesize}
\noindent 
Experiments were conceived and designed by Z.W., T.X., and Q.-F.Y. Measurements and data analysis were performed by Z.W., T.X., with assistance from K.Z., X.L and B.N. Numerical simulations were performed by T.X. with assistance from Z.W. The device was designed by K.Z. and T.X. with assistance from X.L. The device was fabricated and tested by Y.W., with assistance from H.L., J.W., Y.Y. and B.-B.L. The project was supervised by Q.G., Y.-F.X. and Q.-F.Y. All authors participated in preparing the manuscript.

\end{footnotesize}
\medskip

\noindent\textbf{Competing interests}

\begin{footnotesize}
\noindent 
Q.-F.Y., B.N., T.X., Z.W., Y.-F.X. and Q.G. are co-inventors on Chinese patent application CN202510557940.9 concerning the soliton microcomb module.
\end{footnotesize}
\medskip

\noindent\textbf{Data availability}

\begin{footnotesize}
\noindent The data that support the plots within this paper and other findings of this study are available upon publication. 
\end{footnotesize}
\medskip

\noindent\textbf{Code availability}

\begin{footnotesize}
\noindent The codes that support the findings of this study are available upon publication. 
\end{footnotesize}
\medskip

\noindent\textbf{Additional information}

\begin{footnotesize}
\noindent Correspondence and requests for materials should be addressed to Q-F.Y.
\end{footnotesize}

\end{document}


\title{Supplementary information: Integrated soliton microcombs beyond the turnkey limit}

\author{Ze Wang$^{1,*}$, Tianyu Xu$^{1,*}$, Yuanlei Wang$^{1,2*}$, Kaixuan Zhu$^{1,*}$, Xinrui Luo$^{1,}$, Haoyang Luo$^{1}$, Junqi Wang$^{1}$, Bo Ni$^{1,3}$, Yiwen Yang$^{1}$, Qihuang Gong$^{1,3,4}$, Yun-Feng Xiao$^{1,3,4}$, Bei-Bei Li$^{2}$, and Qi-Fan Yang$^{1,3,4\dagger}$\\
$^1$State Key Laboratory for Artificial Microstructure and Mesoscopic Physics and Frontiers Science Center for Nano-optoelectronics, School of Physics, Peking University, Beijing 100871, China\\
$^2$Beijing National Laboratory for Condensed Matter Physics, Institute of Physics, Chinese Academy of Sciences, Beijing 100190, China\\
$^3$Peking University Yangtze Delta Institute of Optoelectronics, Nantong, Jiangsu 226010, China\\
$^4$Collaborative Innovation Center of Extreme Optics, Shanxi University, Taiyuan 030006, China\\
$^{*}$These authors contributed equally to this work.\\
$^{\dagger}$Corresponding author: leonardoyoung@pku.edu.cn}

\date{\today}

\maketitle

\section{Self-injection locking master equations}

We model the self-injection-locked (SIL) system in terms of three fields: the soliton field in the nonlinear resonator (NR), $A_{\mathrm{S}}$; the backscattered field, $A_{\mathrm{B}}$; and the pump laser field, $A_{\mathrm{L}}$. The auxiliary resonator (AR) is treated as producing an effective frequency shift $\delta$ of the NR pump mode relative to the unperturbed NR mode family~\cite{helgason2023surpassing}. The coupled equations of motion are~\cite{kondratiev2017self,godey2014stability,yu2021spontaneous}:
\begin{equation}
       \frac{\partial A_\mathrm{S}}{\partial T}  = -\frac{\kappa_{\mathrm{NR}}}{2}A_\mathrm{S} - i\delta \omega_{\mathrm{c}} A_\mathrm{S} + i\frac{D_{2}}{2}\frac{\partial^2 A_\mathrm{S}}{\partial \phi^2}
       + i g_{\mathrm{NR}}\!\left(\left|A_\mathrm{S}\right|^2+2\left|A_\mathrm{B}\right|^2\right)A_\mathrm{S}
       + i\delta \overline{A_\mathrm{S}}
       + i\beta\frac{\kappa_\mathrm{NR}}{2}A_\mathrm{B}
       + \sqrt{\kappa_{\mathrm{e,NR}}\kappa_{\mathrm{L}}}\,e^{i\phi_{\mathrm B}}A_\mathrm{L},
\end{equation}
\begin{equation}
       \frac{\partial A_\mathrm{B}}{\partial T}  = -\frac{\kappa_{\mathrm{NR}}}{2}A_\mathrm{B} - i\delta \omega_{\mathrm{p}} A_\mathrm{B}
       + i g_{\mathrm{NR}}\!\left(\left|A_\mathrm{B}\right|^2
       + 2\int_{0}^{2\pi} \frac{\left|A_\mathrm{S}\right|^2}{2\pi}\,d\phi\right)A_\mathrm{B}
       + i\beta\frac{\kappa_{\mathrm{NR}}}{2}\overline{A_\mathrm{S}},
\end{equation}
\begin{equation}
       \frac{\partial A_\mathrm{L}}{\partial T}  =
       -\frac{\kappa_{\mathrm{L}}}{2}A_\mathrm{L}
       - i\left(\delta \omega_{\mathrm{p}}-\delta\omega_{\mathrm{L}}\right) A_{\mathrm{L}}
       + \frac{g(|A_{\mathrm{L}}|^2)}{2}(1+i\alpha_{\mathrm{g}})A_{\mathrm{L}}
       + \sqrt{\kappa_{\mathrm{e,NR}}\kappa_{\mathrm{L}}}\,e^{i\phi_{\mathrm{B}}}A_{\mathrm{B}}.
\label{eq:laser}
\end{equation}

\noindent
Here, $T$ is the slow (laboratory) time and $\phi$ is the angular coordinate in the co-rotating frame. $A_\mathrm{S}(T, \phi)$ corresponds to the slowly varying field amplitude of the NR intracavity field; the quantities $\int_{0}^{2\pi} \left|A_\mathrm{S}\right|^2 d\phi/(2\pi)$, $\lvert A_{\mathrm{B}}\rvert^{2}$, and $\lvert A_{\mathrm{L}}\rvert^{2}$ are normalized to intracavity photon numbers. $\overline{A_\mathrm{S}}=\int_{0}^{2\pi} A_\mathrm{S} d\phi/(2\pi)$ is the field amplitude in the NR pump mode. $D_{2}$ is the group velocity dispersion (GVD) of the NR. The total NR decay rate is $\kappa_{\mathrm{NR}} = \kappa_{\mathrm{0,NR}} + \kappa_{\mathrm{e,NR}}$, where $\kappa_{\mathrm{0,NR}}$ is the intrinsic loss rate and $\kappa_{\mathrm{e,NR}}$ is the coupling rate to the bus waveguide. $\kappa_\mathrm{L}$ is the laser mode loss rate. The NR nonlinear coefficient is
$g_{\mathrm{NR}}=\hbar \omega_0^2 c n_2/(n_0^2 V_{\mathrm{eff}})$,
where $V_{\mathrm{eff}}$ is the effective mode volume of NR, $n_0$ is the refractive index, and $n_2$ is the Kerr nonlinear index. $\beta$ is the dimensionless backscattering coefficient (normalized to $\kappa_\mathrm{NR}/2$). The detunings $\delta\omega_\mathrm{p}$ and $\delta\omega_\mathrm{c}$ denote the frequency offsets of the hybrid and cold NR resonances from the SIL laser frequency, with $\delta\omega_\mathrm{c} = \delta\omega_\mathrm{p} + \delta$. The feedback phase $\phi_{\mathrm{B}}$ is the phase delay between the NR and the laser. The detuning $\delta\omega_\mathrm{L}$ is defined relative to the cold laser frequency. The intensity-dependent gain is given by
$g(|A_{\mathrm{L}}|^2)= g_0/(1+|A_\mathrm{L}|^2/|A_\mathrm{L,sat}|^2)$,
where $g_0$ is the small-signal gain, $\lvert A_{\mathrm{L,sat}}\rvert^{2}$ is the saturation photon number, and $\alpha_\mathrm{g}$ is the linewidth-enhancement factor.

Under the assumption that the laser relaxation is fast compared with the cavity dynamics, Eq.~\eqref{eq:laser} can be adiabatically eliminated, leading to an Adler-like description of the SIL dynamics~\cite{shen2020integrated}. For convenience, we introduce normalized variables and recast the equations of motion as:
\begin{equation}
    \frac{\partial \psi_{\mathrm{S}}}{\partial \tau}
    =-(1 + i\zeta)\psi_{\mathrm{S}}
      + i d_{2}\frac{\partial^2 \psi_{\mathrm{S}}}{\partial \phi^2}
      + i\bigl(|\psi_\mathrm{S}|^2+2|\psi_\mathrm{B}|^2\bigr)\psi_\mathrm{S}
      + i\beta\psi_\mathrm{B}
      + i\Delta\overline{\psi_\mathrm{S}}
      + f,
\label{eq:lle_CW}
\end{equation}
\begin{equation}
    \frac{\partial \psi_{\mathrm{B}}}{\partial \tau}
    =-(1 + i\xi)\psi_{\mathrm{B}}
      + i\bigl(|\psi_{\mathrm{B}}|^2+2P\bigr)\psi_{\mathrm{B}}
      + i\beta\overline{\psi_\mathrm{S}},
\label{eq:lle_CCW}
\end{equation}
\begin{equation}
    \xi=\Delta_\mathrm{L}
    + K_0\,\mathrm{Im}\!\left[(1-i\alpha_g)e^{i2\phi_B}\frac{\psi_{\mathrm B}}{f}\right].
\label{eq:lle_inj}
\end{equation}

\noindent
The normalized quantities are defined as:
\[
\tau = \frac{\kappa_{\mathrm{NR}}}{2}T,\quad
\psi_\mathrm{S} = \sqrt{\frac{2g_{\mathrm{NR}}}{\kappa_{\mathrm{NR}}}}\,A_{\mathrm{S}},\quad
\psi_\mathrm{B} = \sqrt{\frac{2g_{\mathrm{NR}}}{\kappa_{\mathrm{NR}}}}\,A_{\mathrm{B}},
\]
\[
P=\int_{0}^{2\pi} \frac{\left|\psi_\mathrm{S}\right|^2}{2\pi}\,d\phi,\quad
\zeta = \frac{2\delta\omega_{\mathrm{c}}}{\kappa_\mathrm{NR}},\quad
\xi = \frac{2\delta\omega_{\mathrm{p}}}{\kappa_\mathrm{NR}},
\]
\[
\Delta_{\mathrm{L}} = \frac{2\delta\omega_{\mathrm{L}}+\kappa_\mathrm{L}\alpha_\mathrm{g}}{\kappa_{\mathrm{NR}}},\quad
\Delta = \frac{2\delta}{\kappa_{\mathrm{NR}}},\quad
d_2= \frac{D_{2}}{\kappa_{\mathrm{NR}}},
\]
\[
f = \frac{2\sqrt{\kappa_\mathrm{e,NR}\kappa_\mathrm{L}}}{\kappa_\mathrm{NR}}
     \sqrt{\frac{2g_{\mathrm{NR}}}{\kappa_{\mathrm{NR}}}}\,A_\mathrm{L}e^{i\phi_\mathrm{B}},\quad
K_0=\frac{4\kappa_\mathrm{e,NR}\kappa_\mathrm{L}}{\kappa_\mathrm{NR}^2}.
\]
In this normalized form, $|f|^2$ is the pump power expressed in units of the parametric-oscillation threshold of the standalone NR (i.e., at $\Delta=0$). The quantities $\xi$, $\zeta$, and $\Delta_{\mathrm L}$ are the normalized pump, comb, and laser detunings, respectively. The normalized mode shift $\Delta$ links the pump and comb detunings via $\zeta=\xi+\Delta$, and $K_{0}$ is the normalized locking bandwidth. From Eq.~\eqref{eq:lle_inj}, the linewidth-enhancement factor $\alpha_{\mathrm g}$ only introduces constant offsets to $\Delta_{\mathrm L}$, $\phi_{\mathrm B}$, and $K_{0}$; we therefore set $\alpha_{\mathrm g}=0$ in the simulations without loss of generality.

The coupled equations~\eqref{eq:lle_CW}–\eqref{eq:lle_inj} are numerically integrated using a split-step Fourier method. Unless stated otherwise, the parameters common to all simulations are $|f|^2=5$, $\beta=0.1$, $K_0=50$, and $d_2=0.02$. For a given mode shift $\Delta$, we sweep $\Delta_{\mathrm{L}}$ and $\phi_{\mathrm{B}}$, injecting quantum noise at the level of half a photon per mode into $\psi_{\mathrm{S}}$, to map the comb-detuning interval that supports turnkey initiation of soliton. A soliton obtained in the turnkey regime is then used as the initial condition to assess soliton persistence in the extended regime.

\begin{figure*}
  \centering
  \includegraphics[width=\linewidth]{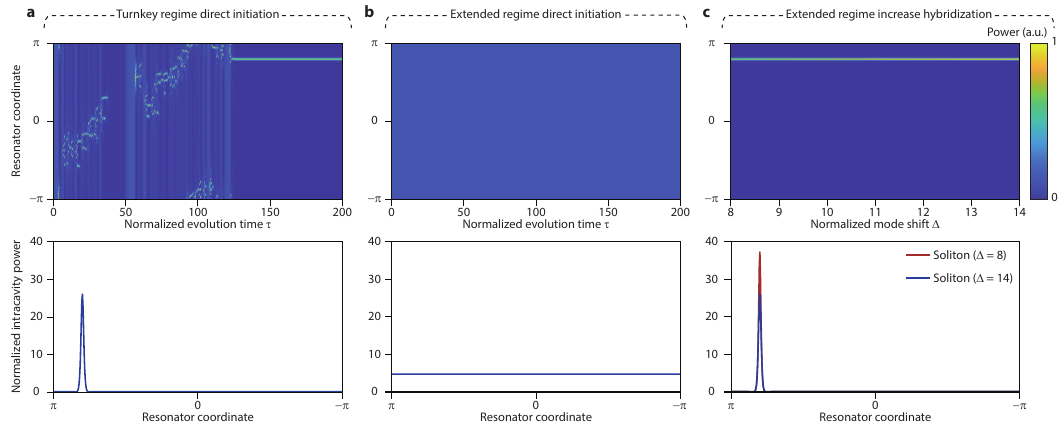}
  \caption{Numerical simulations of soliton dynamics in the turnkey and extended regimes.
  \textbf{a,} Deterministic single-soliton generation when the parametric gain exceed loss. Simulations are seeded with quantum noise corresponding to one-half photon per mode. Parameters: $\phi_\mathrm{B}=0.6\pi$, $\Delta_\mathrm{L}=3.9$, and $\Delta=8$. Upper panel: intracavity power distribution versus normalized evolution time $\tau$ and angular coordinate $\phi$. Lower panel: snapshot of the intracavity power profile at $\tau=200$.
  \textbf{b,} Soliton generation is suppressed when the parametric gain is below loss. Simulations are seeded as in \textbf{a}. Parameters: $\phi_{\mathrm{B}}=0.6\pi$, $\Delta_{\mathrm{L}}=3.9$, and $\Delta=14$. Panels are organized as in \textbf{a}.
  \textbf{c,} A soliton initialized in the turnkey regime transitions into the extended regime as the mode shift is increased from $\Delta=8$ to $\Delta=14$ with all other parameters held fixed. The simulation is seeded with the soliton from \textbf{a}. Upper panel: intracavity power distribution versus normalized mode shift $\Delta$ and angular coordinate $\phi$. Lower panel: snapshots of the intracavity field at $\Delta=8$ (turnkey) and $\Delta=14$ (extended).}
  \label{fig:figS1}
\end{figure*}

Figure~\ref{fig:figS1} summarizes representative simulations in the turnkey and extended regimes. In Fig.~\ref{fig:figS1}a, appropriate choices of $\phi_{\mathrm B}$ and $\Delta_{\mathrm L}$ leads to deterministic turnkey initiation of single-soliton (middle and bottom panels). In Fig.~\ref{fig:figS1}b, a larger mode shift $\Delta$ moves the parametric gain below loss, so sideband oscillation is precluded and only a CW output remains. In Fig.~\ref{fig:figS1}c, seeding with the soliton from Fig.~\ref{fig:figS1}a and increasing $\Delta$ shifts the operating point to larger red detuning while the soliton persists, illustrating the extended regime discussed in the main text.

\section{Parametric gain and operating regimes}

We now analyze the parametric gain to delineate the regions where parametric gain exceed loss and to identify the criterion for deterministic single-soliton initiation. As above, pump-mode hybridization induced by the AR is modeled as an effective frequency shift $\Delta$ of the NR pump mode~\cite{helgason2023surpassing}. When backscattering is weak ($|\beta|\ll1$), the counter-propagating field can be neglected and the soliton field $\psi_{\mathrm S}$ obeys the normalized Lugiato–Lefever equation~\cite{yu2021spontaneous}:
\begin{equation}
\label{eq:LLEparaGain}
\frac{\partial \psi_{\mathrm{S}}}{\partial \tau}
=-(1 + i\zeta)\psi_{\mathrm{S}}
 + i d_{2}\frac{\partial^2 \psi_{\mathrm{S}}}{\partial \phi^2}
 + i|\psi_\mathrm{S}|^2\psi_\mathrm{S}
 + i\Delta\overline{\psi_\mathrm{S}}
 + f,
\end{equation}
\noindent
where all normalized quantities are as defined above. 

Assuming a homogeneous and steady intracavity field, the CW solution of NR pump mode $\psi_{\mathrm{cw}}$ satisfies:
\begin{equation}
\label{eq:CWsolution}
-(1+i\xi)\psi_{\mathrm{cw}} + iP_\mathrm{cw}\psi_\mathrm{cw} + f = 0,
\end{equation}
\noindent
where $P_\mathrm{cw}=|\psi_{\mathrm{cw}}|^{2}$ is the normalized intracavity power in the NR pump mode. To analyze stability, we introduce a small perturbation on the $\mu$th sideband with the ansatz of:
\begin{equation}
\label{eq:perturbation}
\delta\psi(\tau,\phi)
= u\,e^{\lambda\tau+i\mu\phi}
+ v^{*}\,e^{\lambda\tau-i\mu\phi},
\end{equation}
\noindent
which, upon substitution into Eq.~\eqref{eq:LLEparaGain}, yields the eigenvalue $\lambda$ and the parametric gain expression~\cite{godey2014stability}: 
\begin{equation}
\label{eq:paraGain}
G(\mu)=\operatorname{Re}\{\lambda+1\}
= \operatorname{Re}\!\left\{\sqrt{P_{\mathrm{cw}}^{2}
- \bigl(\zeta+d_{2}\mu^{2}-2P_{\mathrm{cw}}\bigr)^{2}}\right\}.
\end{equation}

A necessary condition for soliton initiation is that the parametric gain exceed loss ($G(\mu)\ge 1$). This condition can be visualized in the \((P_{\mathrm{cw}},\zeta)\) plane. The term $\zeta+d_{2}\mu^{2}-2P_{\mathrm{cw}}$ represents the phase mismatch of the $\mu$th sideband. The gain reaches its maximum value $G_{\mathrm{max}}$ when this mismatch is minimized. For anomalous GVD ($d_{2}>0$), two regimes arise. If \(\zeta<2P_{\mathrm{cw}}\), the phase mismatch can vanish at \(\zeta+d_{2}\mu^{2}=2P_{\mathrm{cw}}\), yielding \(G_{\mathrm{max}}=P_{\mathrm{cw}}\) at the preferred sideband index:
\begin{equation}
\label{eq:mu_star}
\mu_\mathrm{*}=\sqrt{\frac{2P_\mathrm{cw}-\zeta}{d_2}}.
\end{equation}
\noindent
If \(\zeta\ge 2P_{\mathrm{cw}}\), the mismatch remains positive and increases monotonically with \(|\mu|\); the gain is then maximized at \(\mu=\pm 1\) (one free spectral range from the pump), favoring deterministic single-soliton formation~\cite{ulanovsynthetic2024}. This leads to the single-soliton criterion:
\begin{equation}
\label{eq:singlesol}
    \zeta\ge 2P_{\mathrm{cw}},
\end{equation}
\noindent
with the corresponding maximum gain:
\begin{equation}
\label{eq:gain}
    G_\mathrm{max}
    =\sqrt{P_{\mathrm{cw}}^{2}
    -\bigl(\zeta+d_{2}-2P_{\mathrm{cw}}\bigr)^{2}}.
\end{equation}

Combining both regimes, the condition for parametric gain to exceed loss becomes:
\begin{equation}
\label{eq:paraGainTot}
\begin{aligned}
    &P_\mathrm{cw} \ge 1, &\quad \zeta\le 2,\\[6pt]
    &\sqrt{P_{\mathrm{cw}}^{2}
    -\bigl(\zeta+d_{2}-2P_{\mathrm{cw}}\bigr)^{2}} \ge 1, &\quad \zeta>2.
\end{aligned}
\end{equation}
\noindent
Equations~\eqref{eq:paraGainTot} and \eqref{eq:singlesol} together delineate the boundary between comb-generating and no-comb domains and partition the comb-generating domain into MI \& multiple-soliton and single-soliton regions.

\begin{figure*}
  \centering
  \includegraphics[width=\linewidth]{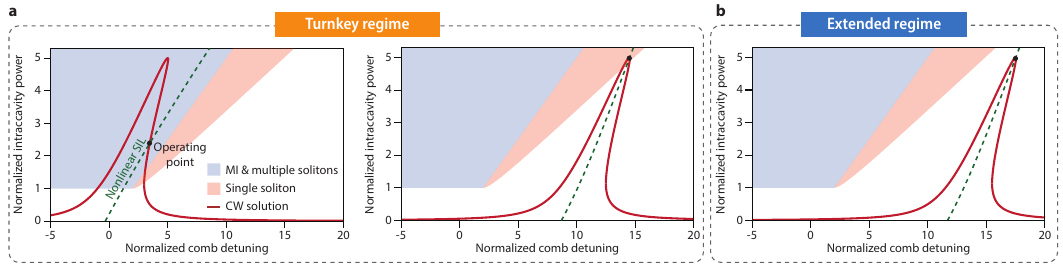}
  \caption{Phase diagram of intracavity power versus comb detuning. The nonlinear SIL trajectory (green dashed line) selects a unique operating point on the CW solution (red curve, $|f|^{2}=5$).
  \textbf{a, } Without mode hybridization (left, $\Delta=0$ and $\phi_\mathrm{B}=0.8\pi$), the operating point occurs at relatively small detuning and typically lies in the MI \& multiple-soliton region. Increasing the hybridization and adjusting the feedback phase (right, $\Delta=9.5$ and $\phi_\mathrm{B}=0.54\pi$) shifts the operating point near the peak of the CW curve and into the single-soliton region.
  \textbf{b, } For larger mode hybridization ($\Delta=12.5$ and $\phi_\mathrm{B}=0.54\pi$), the CW solution overlaps only the no-comb region, so direct soliton initiation is precluded.}
  \label{fig:figS2}
\end{figure*}

Under the infinite locking bandwidth approximation ($K_{0}\to\infty$), the SIL operating point is determined by the feedback phase and intracavity power and is given by the intersection of two constraints: (i) the Kerr-bistable CW pump–power response, described by Eq.~\eqref{eq:CWsolution}, and (ii) the SIL condition, which relates the laser detuning to the intracavity power. The latter can be written as~\cite{shen2020integrated}:
\begin{equation}
\label{eq:Locking}
\bigl(3P_{\mathrm{cw}}^{2}-2\xi\bigr)\cos\varphi
+\bigl(1-2P_{\mathrm{cw}}^{2}+3P_{\mathrm{cw}}\xi-\xi^{2}\bigr)\sin\varphi=0,
\end{equation}
\noindent
where \(\varphi=2\phi_{\mathrm B}-\arctan(\alpha_\mathrm{g})+\pi/2\). When plotted in the \((\zeta,P_{\mathrm{cw}})\) plane, the CW response and SIL condition intersect at the SIL operating point (Fig.~\ref{fig:figS2}). In the absence of mode hybridization (\(\Delta=0\)), this intersection occurs at relatively small comb detuning. Introducing mode hybridization red-shifts the pump resonance, displacing both the CW curve and the SIL curve and moving the operating point to larger detunings. Superimposing the MI \& multiple-soliton, single-soliton, and no-comb regions defined by Eqs.~\eqref{eq:singlesol} and \eqref{eq:paraGainTot} shows that appropriate hybridization enables deterministic, turnkey single-soliton generation (Fig.~\ref{fig:figS2}a), whereas excessive red shift pushes the operating point into the no-comb domain (Fig.~\ref{fig:figS2}b), in agreement with the measurements presented in Fig.~2 of the main text.

\bibliography{SIref.bib}